\documentclass[journal]{IEEEtran}

\IEEEoverridecommandlockouts

\usepackage{cite}
\usepackage{amsmath,amssymb,amsfonts}
\usepackage{algorithmic}
\usepackage{graphicx}
\usepackage{textcomp}
\usepackage{xcolor}
\def\BibTeX{{\rm B\kern-.05em{\sc i\kern-.025em b}\kern-.08em
    T\kern-.1667em\lower.7ex\hbox{E}\kern-.125emX}}
\begin{document}

\title{Index Modulation-Based Flexible Non-Orthogonal Multiple Access
}


		
		    \author{Emre Arslan,~\IEEEmembership{Student Member,~IEEE,} Ali Tugberk Dogukan,~\IEEEmembership{Student Member,~IEEE,}  \\Ertugrul Basar,~\IEEEmembership{Senior Member,~IEEE}
      
\thanks{E. Arslan, A. T. Dogukan and E. Basar are with the Communications Research and Innovation Laboratory (CoreLab),  Department of Electrical and Electronics Engineering, Ko\c{c} University, Sariyer 34450, Istanbul, Turkey. \mbox{Email: earslan18@ku.edu.tr, adogukan18@ku.edu.tr, ebasar@ku.edu.tr}}

\vspace{-0.8cm} }

\maketitle

\begin{abstract}
	Non-orthogonal multiple access (NOMA) is envisioned as an efficient candidate for future communication systems. This letter proposes a novel orthogonal frequency division multiplexing (OFDM) with index modulation (IM)-based NOMA scheme, called OFDM-IM NOMA, for future multi-user communication systems. Inspired by IM and classical OFDM-NOMA, users utilize flexibility by adjusting power allocation factors and subcarrier activation ratios. Our new scheme allows different service users to share available resources as in classical NOMA, more efficiently. It is shown that OFDM-IM NOMA reliably supports a high and low data rate user at the same resources by adjusting their subcarrier activation ratios.
	
\end{abstract}
\vspace{-0.3cm}
	\section{Introduction}
\IEEEPARstart{W}{ireless} communication networks (5G and beyond) demand stringent requirements with diverse users and rich use-cases. For instance, 6G wireless networks are estimated to support up to $10^7$ devices/km$^2$, about ten times the connectivity density when compared to 5G \cite{magazine6G}. To meet these demands, multiple access schemes are being developed to allow multiple users to share the available bandwidth for spectrum efficiency. Recently, non-orthogonal multiple access (NOMA) has been proposed as a promising multiple access candidate, most popularly in the power domain, to enhance spectrum efficiency \cite{6692652}. Power-domain NOMA allows multiple users to share the same time and frequency resources separating them by their different power levels \cite{PDNOMA}. The non-orthogonality in NOMA allows the transmission of multiple signals through superposition coding (SC). Upon reception, successive interference cancellation (SIC) is employed to remove the interference of other users' signals for efficient detection. NOMA additionally introduces flexibility in the power domain and fairness to users with worse channel conditions as well as achieves low-latency. As a result, NOMA is envisioned as a potential candidate for a fair, efficient, low-latency, and flexible multiple access technique allowing wireless connectivity for billions of devices. The literature on NOMA is very rich and many interesting concepts on power domain NOMA have been studied \cite{8457925,li2019spatial,8703780,7887760}. However, state-of-the art NOMA studies usually do not take into consideration of imperfections in SIC detection for bit error rate (BER) analysis and consider both users with the same application demands allowing no flexibility in spectral and energy efficiency. In other words, both users consider the same system parameters in almost all studies. In this paper, we challenge this status quo by providing a flexible waveform design for users with different requirements.

\begin{figure*}[t]
	\centering
	\includegraphics[width=0.8\linewidth,height=5.4cm]{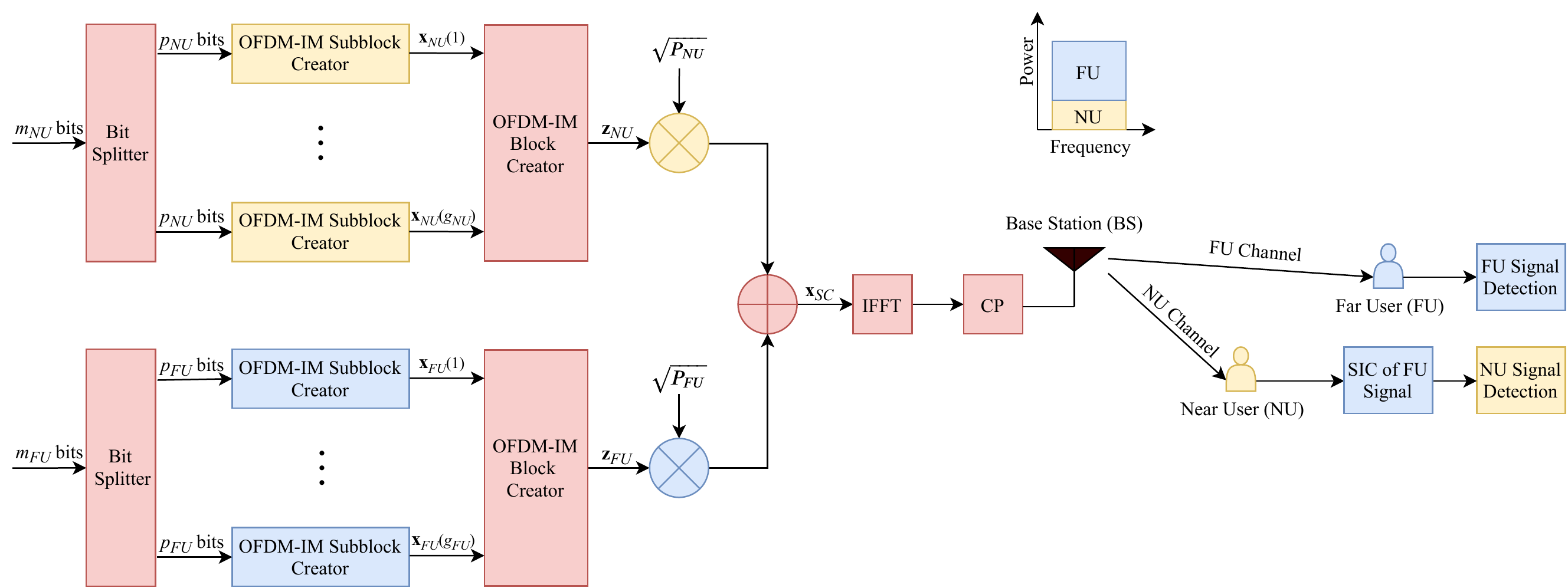}
	\caption{Block diagram of the proposed OFDM-IM NOMA system.}
	\label{Fig. 1}
	\vspace{-0.63cm}
\end{figure*} 

Index modulation (IM) appears as a simple, spectral and energy efficient scheme to transmit additional information bits through the indices of communication systems \cite{IM,basar2017index}. A novel multi-carrier waveform called orthogonal frequency division multiplexing with IM (OFDM-IM) \cite{OFDMIM}, conveys additional information through the indices of subcarriers in the frequency domain along with $M$-ary modulated symbols. This provides an additional degree of freedom by utilizing subcarrier activation ratios where the subcarriers of an OFDM system can be turned on and off according to user applications; therefore, the users may exploit the flexibility of IM. In \cite{8011316}, a NOMA multicarrier index keying (NOMA-MCIK) scheme, which combines IM and NOMA, is proposed. NOMA-MCIK applies IM to subblocks which consist of a certain number of subcarriers and transmits the same modulated symbol through the active subcarriers. Using the subcarrier activation feature of OFDM-IM and flexibility of NOMA in the power domain to multiplex users, it is possible to construct a novel and flexible waveform to ensure  high spectrum efficiency depending different user requirements. 

The aim of this paper is to propose a flexible scheme by incorporating IM to the subcarriers of OFDM-NOMA users, where the users exploit the same or different number of active subcarriers. The flexibility provided from both NOMA by adjusting the power allocation factors and OFDM-IM with adjustable subcarrier activation ratios, is taken advantage of, to provide a promising and alternative waveform called OFDM-IM NOMA. Different from NOMA-MCIK, OFDM-IM NOMA applies IM to the subcarriers in a subblock and transmits different symbols over active subcarriers to increase spectral efficiency. The proposed scheme is based on the simultaneous transmission of two users' OFDM-IM signals with different power levels where the information is conveyed through both the active subcarrier indices and $M$-ary modulated symbols. In this setup, a main user (the near user) that needs higher data rates, and multiple far users that have low data rate requirements, i.e., Internet-of-things (IoT) devices, share the same spectrum. Classical OFDM-NOMA provides the same amount of subcarriers for both users while assigning different power levels \cite{8778559}. Using IM, the proposed scheme activates most of the subcarriers for the near user and a fewer number of ones for the IoT-based far users. Since in OFDM-IM NOMA the far user has a fewer number of activated subcarriers, it may exploit lower order modulations without SIC decoding. As a result, the far user utilizes only a small amount of the spectrum according to its modest requirements while enjoying a reliable and energy efficient transmission. Meanwhile, the near user may continue to use the spectrum with a higher data rate benefiting from the flexibility provided by OFDM-IM. At the near user, SIC is used to eliminate the high power far user interference. Benefits of the proposed scheme to classical OFDM-NOMA include higher flexibility with different subcarrier activation ratios which allow adjustable spectral efficiencies and controllable inter-user interference, higher energy efficiency, improved reliability and higher spectrum efficiency. For the proposed scheme, an optimum power allocation search is performed and theoretical error performance is investigated to confirm our computer simulations. Taking a different approach from the current NOMA literature, this paper focuses particularly on BER analyses and flexibility in the waveform domain for the case of two users rather than sum-rate and spectral efficiency analyses.

The rest of the paper can summarized as follows. In Section II, the system model of OFDM-IM NOMA is presented. In Section III, the theoretical error performance of the proposed scheme is investigated along with the optimum power allocation method. Our computer simulation results are presented in Section IV. Finally, Section V concludes the paper.
\vspace{-0.3cm}
\section{System Model}

In this section, the system model of the proposed OFDM-IM NOMA scheme is given. First the transmitter architecture is presented, then the receiver model is explained.
\vspace{-0.4cm}
\subsection{Transmitter Side}
Let us consider an OFDM-based downlink (DL) NOMA system operating under frequency-selective Rayleigh fading channels. The base-station (BS) generates OFDM-IM signals to serve $U$ users independently through the same frequency and time slots by SC. The BS implements SC by allocating different power levels to multiple users and transmitting a composite signal. This paper considers a scenario with only two users $(U=2)$ as shown in Fig. 1.  For each user, a total of $m_u$ bits are transmitted, where $u \in \{NU,FU\}$, and NU and FU denote near and far users, respectively. These $m_u$ bits are split into $g_u$ groups, each containing $p_u=m_u/g_u$ bits. Since the same procedures are applied for the creation of each subblock, it is sufficient to consider the $\xi^{\text{th}}$ subblock ${\mathbf{x}}_{u}(\xi) \in \mathbb{C}^{n_u \times 1}$ of any user, $\xi \in \{1,\cdots,g_u\}$. The $p_u$ bits are mapped into subblocks with sizes $n_u$, where $n_u=N/g_u$ and $N$ is the number of total subcarriers, that is, the size of fast fourier transform (FFT). For OFDM-IM subbblock creation, the $p_u$ bits are separated into two parts. First, $p_{u,1}=\lfloor\log_2\binom{n_u}{k_u}\rfloor$ bits determine $k_u$ active subcarriers in a subblock using a look-up table or the combinatorial method \cite{OFDMIM}, and produces the active subcarrier index set as $\mathcal{I}_u(\xi)=\{i_{u,1}(\xi),\cdots,i_{u,k_u}(\xi)\}$ where $i_{u,\lambda}(\xi)\in \{1,\cdots,n_u\}$ and $\lambda=1,\cdots,k_u$. Then, the $p_{u,2}=k_u\log_2(M_u)$ bits determine the vector of modulated symbols $\mathbf{q}_u(\xi)=[q_{u,1}(\xi),\cdots,q_{u,k_u}(\xi)]$, $q_{u,\lambda}(\xi)\in \mathcal{Q}_u$, conveyed through the active subcarriers, $\mathcal{Q}_u$ is the complex signal constellation of user $u$ with the size of $M_u$ and is normalized to unit average power. Consequently, a total of 
\begin{equation}
p_u=p_{u,1}+p_{u,2}=\lfloor\log_2\binom{n_u}{k_u}\rfloor+k_u\log_2(M_u)
\end{equation}
bits are transmitted via each subblock. The $N\times1$ OFDM-IM signal block of each user is given by ${\mathbf{z}}_u=[\mathbf{x}_{u}^{\mathrm{T}}(1)\,\mathbf{x}_{u}^{\mathrm{T}}(2)\,\cdots\,\mathbf{x}_{u}^{\mathrm{T}}(g_u)]^{\mathrm{T}}=[x_u(1)\,x_u(2)\,\cdots\, x_u(N)]^{\mathrm{T}}$ and it is created by considering $\mathcal{I}_u(\xi)$ and $\mathbf{q}_u(\xi)$, subblock by subblock, where $x_u(\varphi)\in \{0,\mathcal{Q}_u\}$, $\varphi=1,\cdots,N$ and  $(.)^{\mathrm{T}}$ denotes transposition. Then, SC is applied in the frequency domain to obtain the overall transmission vector as:

\begin{equation}
{\mathbf{x}}_{SC}=\sqrt{\alpha P_{BS}} \mathbf{z}_{NU}+\sqrt{(1-\alpha) P_{BS}} \mathbf{z}_{FU}
\end{equation}
where $P_{BS}$ and $\alpha$ is the total transmit power of the BS (per subcarrier) and the power allocation factor, respectively. Here, we consider $P_{BS}=1$ and $0\leq \alpha \leq 1 $. The average power of the NU and the FU per subcarrier are $P_{NU}=\alpha P_{BS}$ and $P_{FU}=(1-\alpha) P_{BS}$, respectively. 

After this point, the procedures as in OFDM are applied. A block-type interleaver is employed to eliminate the correlation effect of the channel in a subblock as in \cite{6841601}. The inverse fast fourier transform (IFFT) is applied to obtain the time domain OFDM block $\mathbf{x}_{TI}$:
\begin{equation}
{\mathbf{x}}_{TI}=\mathrm{IFFT}\big\{{\mathbf{x}}_{SC}\big\} = \begin{bmatrix} X(1) & X(2) & \cdots & X(N) \end{bmatrix}^\mathrm{T}.
\end{equation}
\noindent

After the IFFT operation, a cyclic prefix (CP) of length $C$ samples $[ X(N-C+1)  \cdots  X(N-1)X(N) ]^{\mathrm{T}}$ is added to the beginning of the OFDM block. Once parallel to serial (P/S) and digital/analog conversions are applied, the signal is transmitted over a frequency-selective Rayleigh fading channel, which can be represented by the channel impulse response (CIR) coefficients
\begin{equation}
\mathbf{t}=\begin{bmatrix} t(1) & \cdots & t(v) \end{bmatrix}^\mathrm{T}
\end{equation}
where $t(\zeta),\zeta=1,\dots,v$ are circularly symmetric complex Gaussian random variables with the $\mathcal{C}\mathcal{N}(0,\frac{1}{v})$ distribution. With the assumption that the channel remains constant during transmission of an OFDM block and the CP length $C$ is larger than $v$, the received vector of two users can be given by 
\begin{equation}
\mathbf{y}_{u}=[y_{u}(1)\cdots y_{u}(N)]^{\mathrm{T}}=\mathbf{H}_{u}\mathbf{x}_{SC}+\mathbf{w}_{u}
\end{equation}
where $\mathbf{H}_{u}=\mathrm{diag}(\mathbf{h}_{u})=\mathrm{diag}([h_{u}(1)\,\cdots\, h_{u}(N)]^{\mathrm{T}})$ and $\mathbf{w}=[w_{u}(1)\,\cdots\, w_{u}(N)]^{\mathrm{T}}$ are the channel matrix and the noise vector for the two users in the frequency domain, respectively, and $\mathrm{diag}(\cdot)$ converts a vector into a diagonal matrix. The distributions of $h_{u}(\varphi)$ and $w_{u}(\varphi)$ are $\mathcal{C}\mathcal{N}(0,\sigma^2_u)$ and $\mathcal{C}\mathcal{N}(0,N_{0})$, respectively, where $N_{0}$ is the noise variance in the frequency domain, which is equal to the noise variance in the time domain. The signal-to-noise ratio (SNR) is defined as $1/N_{0}$. The spectral efficiency of the users is given by
\begin{equation}
\vspace{-0.15cm}
\eta_u =m_u/(N+C) \hspace{0.2cm} \text{[bits/s/Hz]}.  	
\end{equation}

\vspace{-0.3cm}
\subsection{Receiver Side}
In this subsection, the receiver structure is presented for both users. It is assumed that the channel gain for the NU is higher than FU $(\sigma^2_{NU} \geq \sigma^2_{FU})$ due to the distance difference between them, which is a valid assumption in NOMA-based multi-user systems. At both receivers, after obtaining $\mathbf{y}_{u}$, a block-type de-interleaver is employed. For both users, we define the set of all possible subblock realizations as $\mathcal{B}_u=\{\mathbf{b}_{u,1},\mathbf{b}_{u,2},\cdots,\mathbf{b}_{u,G_u}\}$ and the total number of possible subblocks as ${G}_{u}=2^{p_u}$, $u\in\{NU,FU\}$.
\subsubsection{Near User (NU)}
Since the signal power of the NU is less than the FU $(\alpha<0.5)$, first the FU signal is decoded through SIC. We can decode the $\xi^{\mathrm{th}}$ subblock of the FU using the set  $\mathcal{B}_{FU}$. Finally, maximum-likelihood (ML)-based detection rule for SIC operation can be employed as
\vspace{-0.2cm}
\begin{equation}
\hat{\mathbf{x}}_{FU,\text{SIC}}(\xi)=\underset{\mathbf{b}_{FU}\in\mathcal{B}_{FU}}{\arg\: \min}{ \left \|\Bar{\mathbf{y}}_{NU}(\xi)-\sqrt{P_{FU}}\Bar{\mathbf{H}}_{NU}(\xi) \mathbf{b}_{FU} \right \|}^2
\end{equation}

where $\Bar{\mathbf{y}}_{NU}(\xi)=[y_{NU}(n_{FU}(\xi-1)+1)\cdots y_{NU}( n_{FU} \xi)]^{\mathrm{T}}\in \mathbb{C}^{n_{FU} \times 1}$ and $\Bar{\mathbf{H}}_{NU}(\xi)=\mathrm{diag}([h_{NU}(n_{FU}(\xi-1)+1)\cdots h_{NU}( n_{FU} \xi)])\in \mathbb{C}^{n_{FU} \times n_{FU}}$, are the received signal vector and the equivalent channel matrix of the FU at the NU for the $\xi^{\mathrm{th}}$ subblock, respectively. Then, the decoded FU block is re-constructed as $\hat{\mathbf{z}}_{FU}=[\hat{\mathbf{x}}_{FU,\text{SIC}}(1) \dots {\hat{\mathbf{x}}}_{FU,\text{SIC}}(g_{FU})]^T \in \mathbb{C}^{N \times 1}$ and subtract it from the overall received OFDM block as follows:
\begin{equation}
{\mathbf{r}}_{NU}=\mathbf{y}_{NU}-\hat{\mathbf{z}}_{FU}=[r_{NU}(1)\cdots r_{NU}(N)]^{\mathrm{T}}. 
\end{equation}
Hence, the interference of the FU signal on the NU signal is eliminated. Finally, the NU signal is decoded by ML-based detection rule with the NU subblock set $\mathcal{B}_{NU}$: 
\begin{equation}
\hat{\mathbf{x}}_{NU}(\xi)=\underset{\mathbf{b}_{NU}\in\mathcal{B}_{NU}}{\arg\: \min}{ \left \|\Bar{\mathbf{r}}_{NU}(\xi)-\sqrt{P_{NU}}\Tilde{\mathbf{H}}_{NU}(\xi) \mathbf{b}_{NU} \right \|}^2
\end{equation}
where $\Bar{\mathbf{r}}_{NU}(\xi)=[r_{NU}(n_{NU}(\xi-1)+1)\cdots r_{NU}( n_1 \xi)]^{\mathrm{T}}\in \mathbb{C}^{n_{NU} \times 1}$ and $\Tilde{\mathbf{H}}_{NU}(\xi)=\mathrm{diag}([h_{NU}(n_{NU}(\xi-1)+1)\cdots h_{NU}( n_{NU} \xi)])\in \mathbb{C}^{n_{NU} \times n_{NU}}$ are the interference eliminated signal vector and the channel matrix corresponding to $\xi^{\mathrm{th}}$ subblock for the NU. Note that for $n_{NU}=n_{FU}$, which means that the same subblock size is used for both users, we have $\Bar{\mathbf{H}}_{NU}(\xi)=\Tilde{\mathbf{H}}_{NU}(\xi)$  .

\subsubsection{Far User (FU)}
Since the signal power of the FU is higher than the NU $(\alpha<0.5)$, it is assumed that the interference of the NU signal on the FU signal is negligible. Therefore, we can decode the $\xi^{\mathrm{th}}$ subblock of the FU directly using the FU set $\mathcal{B}_{FU}$. As a result, ML-based detection rule can be employed as:
\begin{equation}
\hat{\mathbf{x}}_{FU}(\xi)=\underset{\mathbf{b}_{FU}\in\mathcal{B}_{FU}}{\arg\: \min}{ \left \|\Bar{\mathbf{y}}_{FU}(\xi)-\sqrt{P_{FU}}\Bar{\mathbf{H}}_{FU}(\xi) \mathbf{b}_{FU} \right \|}^2
\end{equation}
where $\Bar{\mathbf{y}}_{FU}(\xi)=[y_{FU}(n_{FU}(\xi-1)+1)\cdots y_{FU}(n_{FU}\xi)]^{\mathrm{T}}\in \mathbb{C}^{n_{FU} \times 1}$ and $\Bar{\mathbf{H}}_{FU}(\xi)=\mathrm{diag}([h_{FU}(n_{FU}(\xi-1)+1)\cdots h_{FU}( n_{FU}\xi)]^{\mathrm{T}})\in \mathbb{C}^{n_{FU} \times n_{FU}}$, are the received signal vector and the channel matrix for the $\xi^{\mathrm{th}}$ subblock, respectively. 
\vspace{-0.5cm}
\section{Performance Analysis}	
\subsection{Theoretical BER derivation}
In this subsection, a theoretical error probability analysis is provided and an average bit error probability (ABEP) expression is obtained for both the NU and the FU.  It is sufficient to investigate the pairwise error probability (PEP) events of a single subblock to derive the performance of OFDM-IM NOMA since the error performance is identical for all subblocks. Thus, we remove the subblock index $\xi$ in the sequel for simplicity.

From $(5)$, assume that $\mathbf{B}_u=\mathrm{diag}(\mathbf{b}_u)$ is transmitted and $\hat{\mathbf{B}}_u=\mathrm{diag}(\hat{\mathbf{b}}_u)$ is erroneously detected. In this case, the unconditional PEP (UPEP) can be given by\cite{CIOFDMIM}
\begin{equation}
\text{Pr}(\mathbf{B}_u \rightarrow \hat{\mathbf{B}}_u)
=\frac{1}{\pi}\int_{0}^{\pi/2} M_{\delta}\Big(-\frac{\sigma^2_u}{4N_0\sin^2{\theta}}\Big)d\theta.
\end{equation}
where $\delta=\|(\mathbf{B}_u-\hat{\mathbf{B}}_u) \Bar{\mathbf{h}}_{u}\|^2$ and $\Bar{\mathbf{h}}_{u} \in \mathbb{C}^{n_{u} \times n_{u}}$ is the channel vector corresponding to a subblock. Expressing $\delta$ in the quadratic form as $\delta=\Bar{\mathbf{h}}_{u}^{\mathrm{H}}\mathbf{Q}\Bar{\mathbf{h}}_{u}$, where $\mathbf{Q}=(\mathbf{B}_u-\hat{\mathbf{B}}_u)^{\mathrm{H}}(\mathbf{B}_u-\hat{\mathbf{B}}_u)$ and $(.)^{\mathrm{H}}$ denotes Hermitian transposition, its moment generating function can be calculated from \cite{BOOK} as $M_{\delta}(t)=[\mathrm{det}(\mathbf{I}_N-t\mathbf{Q})]^{-1}=[\prod_{i_1}^{q}(1-t\lambda_i)]^{-1}$, where $q=\mathrm{rank}(\mathbf{Q})$ with non-zero eigenvalues of $\mathbf{Q}$ being $\lambda_i,i=1,2,\cdots,q$ for $q\leq N$ and $\mathbf{I}_N$ is an $N\times N$ identity matrix. Substituting $M_{\delta}(t)$ in $(10)$, we obtain
\begin{equation}
\text{Pr}(\mathbf{B}_u \rightarrow \hat{\mathbf{B}}_u)
=\frac{1}{\pi}\int_{0}^{\pi/2} \prod_{i=1}^{q} \Bigg( \frac{\sin^2{\theta}}{\sin^2{\theta}+\frac{\sigma^2_u\lambda_i}{4N_0}} \Bigg)d\theta
\end{equation}
which has a closed form solution in Appendix$(5.\text{A})$ of \cite{BOOK}. After obtaining UPEP of each user, the interference of the FU on the NU is ignored, and ABEP of the NU can be approximated as:
\begin{align}
&P_{b,NU}\approx 0.5P_{s,FU}+    (1-P_{s,FU})\times \nonumber \\& \frac{\sqrt{P_{NU}}}{p_{NU} G_{NU}}  \sum_{\mathbf{B}_{NU}}\sum_{\hat{\mathbf{B}}_{NU}} \text{Pr}(\mathbf{B}_{NU} \rightarrow \hat{\mathbf{B}}_{NU}) \times e(\mathbf{B}_{NU},\hat{\mathbf{B}}_{NU})
\end{align}
where 
\begin{equation}
P_{s,FU}=\frac{\sqrt{P_{FU}}}{ G_{FU}}\sum_{\mathbf{B}_{FU}}\sum_{\hat{\mathbf{B}}_{FU}}\text{Pr}(\mathbf{B}_{FU} \rightarrow \hat{\mathbf{B}}_{FU})
\end{equation}
is the symbol error probability (SEP) of the FU and $e(\mathbf{B}_{u},\hat{\mathbf{B}}_{u})$ stands for the number of bit errors for the corresponding pairwise error event $(\mathbf{B}_u \rightarrow \hat{\mathbf{B}}_u)$. The interference of the NU on the FU signal is neglected, therefore, ABEP of the FU can be lower bounded by
\begin{align}
&P_{b,FU}\geq \nonumber \\  & \frac{\sqrt{P_{FU}}}{p_{FU} G_{FU}}\sum_{\mathbf{B}_{FU}}\sum_{\hat{\mathbf{B}}_{FU}} \text{Pr}(\mathbf{B}_{FU} \rightarrow \hat{\mathbf{B}}_{FU})\times e(\mathbf{B}_{FU},\hat{\mathbf{B}}_{FU}).
\end{align}
As will be shown in next section, (13) and (15) can be used to predict the BER of the NU and the FU, respectively.
\vspace{-0.4cm}
\subsection{Optimum Power Allocation}
\begin{figure}[t]
	\centering
	\includegraphics[width=0.9\columnwidth,height=5.8cm]{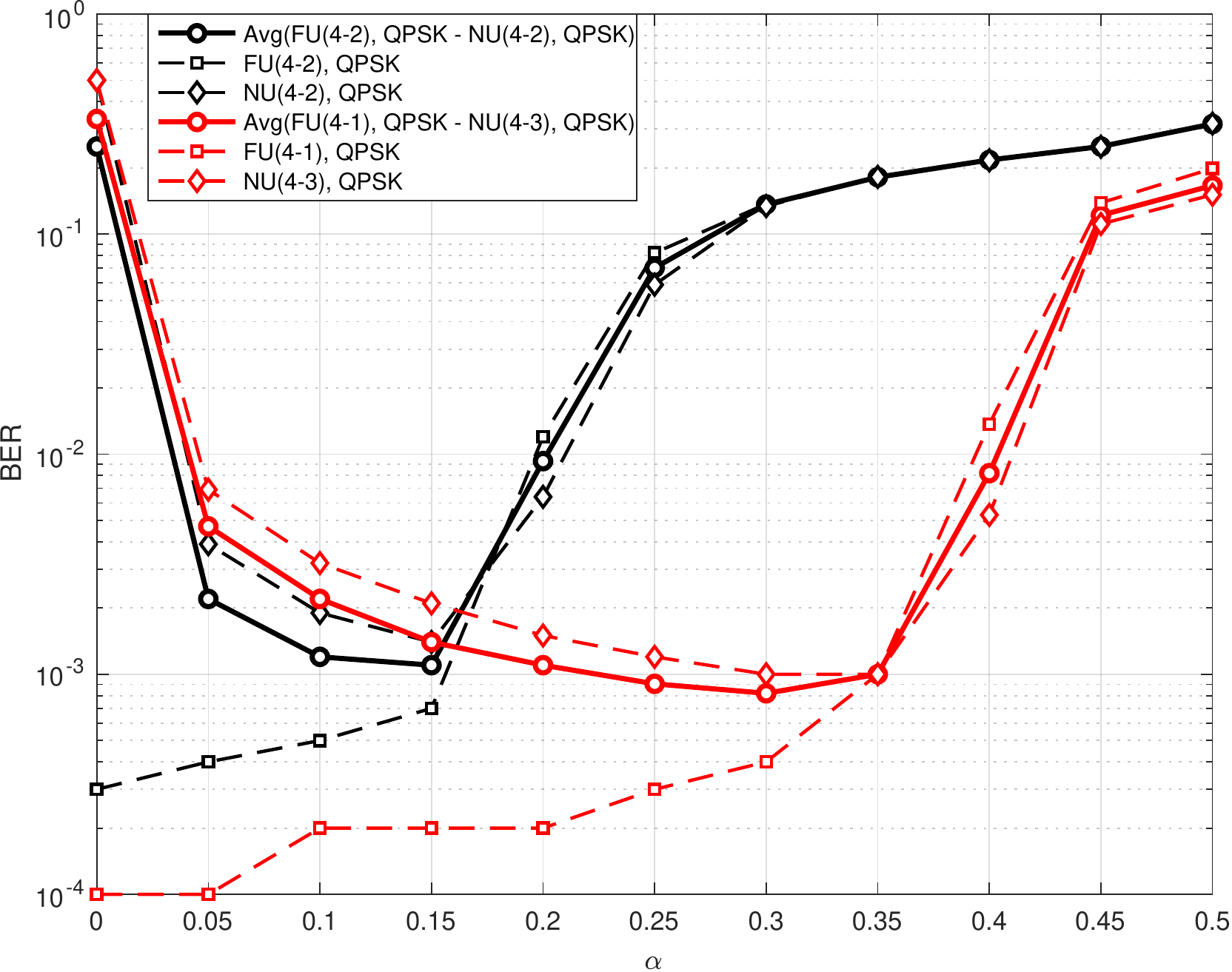}
	\vspace{-0.3cm}
	\caption{Optimum $\alpha$ search from $0$ to $0.5$ with intervals of $0.05$ for a fixed SNR of $30$ dB.}
	\label{Fig. 5}
	\vspace{-0.4cm}
\end{figure}  

Current NOMA literature mostly considers a fixed power allocation without any search method. To obtain better error performance, in this subsection, the power allocation factor $\alpha$ is determined according to the average BER of both users \footnote{Another criteria rather than average BER may be considered for determining the optimum $\alpha$. For example, the error performance of FU can be prioritized.} in the presence of SIC errors and interference using Monte Carlo simulations. For different $k_{NU}$ and $k_{FU}$ values, the performance of the users may vary, therefore, the average performance of the NU and FU is considered: 
\begin{equation}
\text{Avg}_{\text{BER}}=\frac{p_{FU}\text{BER}_{FU}+p_{NU}\text{BER}_{NU}}{p_{FU}+p_{NU}}
\end{equation}
where $\text{BER}_{FU}$ and $\text{BER}_{NU}$ stand for the bit error rate of the far user and the near user, respectively.
As shown in Fig. 2, the average BER of both users for two cases are simulated at a fixed SNR value of $30$ dB for $\alpha$ values varying from $0$ to $0.5$ with an increasing interval of $0.05$. Then the $\alpha$ value that provides the minimum average BER is selected. Fig. 2 shows our power allocation search for two different configurations of the proposed scheme and also consists of the individual BER of both users for each case. We refer to $k$ active out of $n$ subcarriers as $u(n_u,k_u)$, $u\in\{NU,FU\}$. It is observed that different configurations may provide different optimal power allocation factors where for FU(4-2)-NU(4-2) and FU(4-1)-NU(4-3) the optimum $\alpha$ is obtained as $0.15$ and $0.3$, respectively. As the activation ratio of user $u$ is increased more power needs to be allocated to user $u$ and vice-versa as seen in Fig. 2. Note that theoretical derivations of (13) and (15) cannot be used to obtain the optimum $\alpha$ since they do not consider the interference of the NU to the FU.
\vspace{-0.2cm}
\section{Simulation Results}

\begin{figure}[t]
	\centering
	\includegraphics[width=1\columnwidth,height=6.2cm]{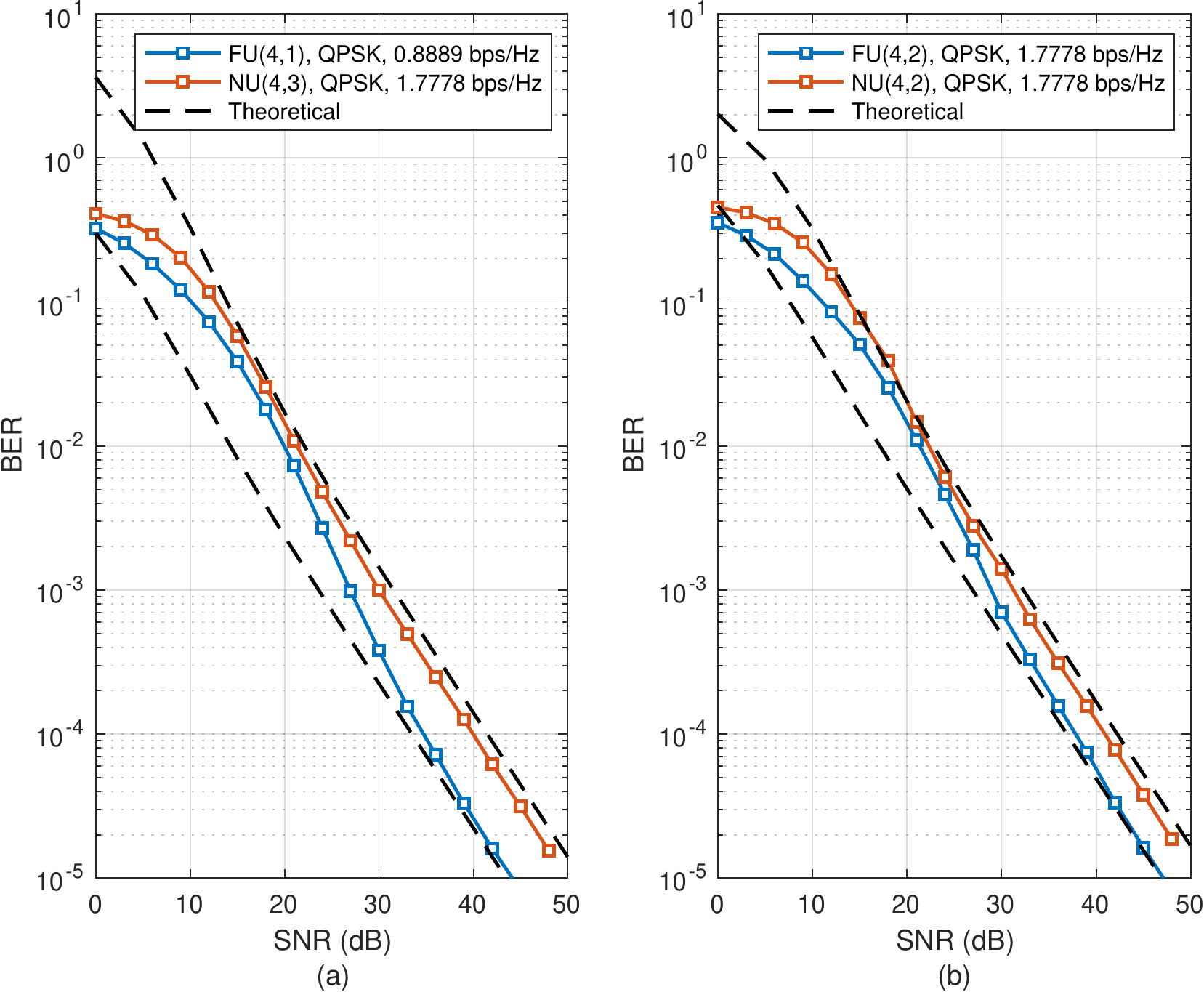}
	\caption{Error performance of OFDM NOMA-IM with $n=4$ for (a) $k_{FU}=1$, $k_{NU}=3$ and (b) $k_{FU}=2$, $k_{NU}=2$.}
	\label{Fig. 5}
	\vspace{-0.2cm}
\end{figure}  

	

	

In this section, computer simulation results are presented for the OFDM-IM NOMA scheme in the presence of Rayleigh fading channels and perfect channel state information at the users. The system parameters are assumed to be: $N=128$, $C=16$, $\sigma^2_{NU}=0$ dB, $\sigma^2_{FU}=-3$ dB and $v=10$. Computer simulation results show the BER performance of these schemes using Monte Carlo simulations. For all simulations, block interleaving of size of $n_u \times g_u$, imperfect SIC detection and a fixed subblock size of $n_{NU}=n_{FU}=4$ is considered. In our simulations, the same subblock size is chosen for ease of presentation, however a generalization is possible. Optimum power allocation factor $\alpha$ is obtained by simulating the averaged BER values from $\alpha=0$ to $\alpha=0.5$ with an increasing interval of $0.05$ for a fixed SNR of $30$ dB and choosing the power allocation factor with the minimum average BER as in Fig. 2. 

Fig. 3 shows the simulation and theoretical results of the proposed OFDM-IM NOMA scheme for two different setups. In Fig. 3(a), $k_{NU}=3$, $k_{FU}=1$, $M_{NU}=4$ and $M_{FU}=4$ are considered. Here, the NU (main user) has a higher spectral efficiency, while the FU (IoT user) has a lower spectral efficiency. It can be seen that the FU exhibits a better BER performance due to less interference along with a lower order modulation and a higher power allocated to it. The use of a lower order modulation, no need for SIC and a better error performance makes the FU more suitable for IoT applications because it can also be more energy efficient due to a lower complex detector architecture. In other words, this user would just need to communicate reliably at low data rates, for example, to transmit sensor data. BER performance of the NU is worse because of the employment of a higher modulation order and remaining interference of the FU during the SIC process along with its lower allocated power. The theoretical curve of the FU obtained by $(15)$ acts as a lower bound due to an assumption of no interference from the NU in the calculation. The theoretical BER curve of the NU is obtained by $(13)$ and is an approximate one because we do not take into consideration of the NU interference when decoding the FU first in the SIC process. From Fig. 3(b), it is also confirmed that the theoretical results match the simulation results with different set of parameters: $k_{NU}=2$, $k_{FU}=2$, $M_{NU}=4$ and $M_{FU}=4$. 

Fig. 4 shows the comparison of classical OFDM-NOMA and OFDM-IM NOMA with varying parameters. All NUs and FUs have a spectral efficiency value of $1.7778$ and $0.8889$ bps/Hz, respectively. It can be seen that along with the main benefits of flexibility of the user applications and high spectral efficiency, we additionally obtain a better error performance for the FU and, most of the time, for the NU. The FU achieves a better error performance due to characteristics of IM as in OFDM-IM. Due to the employment of higher modulation orders, lower power allocation and significant interference at SIC decoding, the NU error performance may be worse at certain configurations. Another reason that negatively affects the error performance is that the interference of the other user harms the OFDM-IM decoding process. This is because in classical OFDM-IM, the inactive subcarriers are always zero, but in OFDM-IM NOMA, there is a possibility they may not be completely zero and an active subcarrier of the other user may act as interference to the main signal. However, as in the case of FU$(4,1)$-NU$(4,3)$, even the NU exhibits a better BER performance than classical OFDM-NOMA since the probability of an active subcarrier of the FU overlapping with an empty subcarrier of the NU is dramatically less than in the case of FU$(4,2)$-NU$(4,2)$.
\begin{figure}[t]
	\centering
	\includegraphics[width=0.80\columnwidth,height=6.6cm]{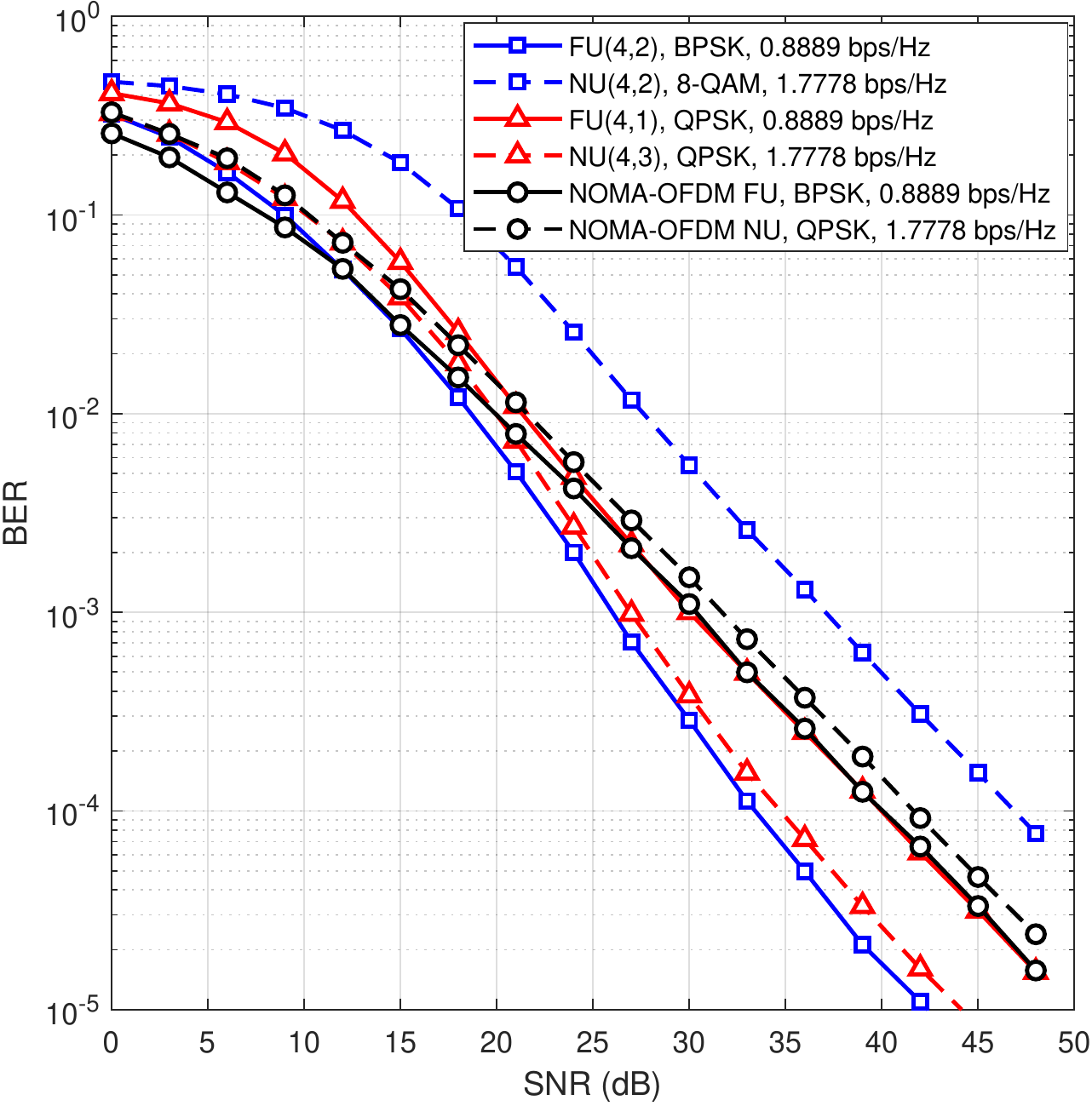}
	\vspace{-0.3cm}
	\caption{Error performance of OFDM NOMA-IM with different configurations compared to OFDM-NOMA.}
	
	\label{Fig. 5}
	\vspace{-0.5cm}
\end{figure} 
\vspace{-0.3cm}
\section{Conclusion}		
In this paper, OFDM-IM NOMA has been presented as a novel and flexible NOMA scheme for different application users with varying subcarrier activation ratios. The proposed scheme allows a main user with a higher data rate and less data rate user such as an IoT device to share the spectrum in a more fair manner. System performance has been shown in terms of BER and imperfect SIC has been considered to investigate a more realistic NOMA scenario. OFDM-IM NOMA has been shown to provide a better BER along with its flexibility advantages. Future extensions may include research for more effective ways to reduce interference between users, enabling more number of IoT users to work with the main user instead of one. An additional improvement may be an advanced optimum power allocation method and the development of an LLR detection method with suitable thresholds to determine the activation probabilities of subcarriers in the subblock for each user.
\vspace{-0.4cm}
\bibliographystyle{IEEEtran}
\bibliography{IEEEabrv,references}

\end{document}